# Operando Control of Skyrmion Density in a Lorentz Transmission Electron Microscope with Current Pulses

Albert M. Park,[1,a] Zhen Chen,[1,a] Xiyue S. Zhang,[1] Lijun Zhu,[1] David A. Muller,[1,2] Gregory D. Fuchs[1,2,b]

[1] School of Applied and Engineering Physics, Cornell University, Ithaca, New York 14853, USA

[2] Kavli Institute at Cornell for Nanoscale Science, Ithaca, New York 14853, USA

ABSTRACT Magnetic skyrmions hold promise for spintronic devices. To explore the dynamical properties of skyrmions in devices, a nanoscale method to image spin textures in response to a stimulus is essential. Here, we apply a technique for operando electrical current pulsing of chiral magnetic devices in a Lorentz transmission electron microscope. In ferromagnetic multilayers with interfacial Dzyaloshinskii-Moriya interaction (DMI), we study the creation and annihilation of skyrmions localized by point-like pinning sites due to defects. Using a combination of experimental and micromagnetic techniques, we establish a thermal contribution for the creation and annihilation of skyrmions in our study. Our work reveals a mechanism for controlling skyrmion density, which enables an examination of skyrmion magnetic field stability as a function of density. We find that high-density skyrmion states are more stable than low-density states or isolated skyrmions resisting annihilation over a magnetic field range that increases monotonically with density.

---

[a] AMP and ZC contributed equally to this work.

[b] Electronic mail: gdf9@cornell.edu

Magnetic skyrmions are topological spin textures that could potentially store and process information with non-volatility, high speed, and low power consumption.[1–6] The recent discovery of homochiral Néel skyrmions in asymmetric ferromagnetic multilayers with interfacial DMI further enhances the potential of skyrmion-based devices because they are stable at room temperature, and because the films supporting this type of skyrmion are compatible with modern nanofabrication processes.[7–13] In addition, manipulation of skyrmions with electric current is possible using spin-orbit torque at ferromagnet/heavy metal interfaces,[7,9,13–20] which enables skyrmion speeds exceeding 100 m/s on a racetrack.[7,17,19]

Pioneering skyrmion research used micro-scale skyrmions that can be tracked with magneto-optical microscopy.[2,9,13,20–22] However, the push for high-density information processing has led to the use of skyrmions at or below the 100 nm length scale, which requires higher resolution imaging techniques. Lorentz transmission electron microscopy (LTEM), with spatial resolution on the order of a few nanometers and time-resolution of sub-seconds,[23] stands out as an excellent option to study both the magnetic structure and evolution of skyrmions. Although some of the most urgent questions regarding skyrmion dynamics involve control using electric current, few studies report operando current control in an LTEM setting. Furthermore, most of these operando biasing studies are on single-crystalline materials, thinned and micropatterned by focused ion beam (FIB).[24–26] In the case of ferromagnetic multilayers with interfacial DMI, only millimeter-scale sample areas with a DC bias have been reported.[27] Almost contemporary to this work, current pulses have been applied to achieve high density skyrmions in a synthetic antiferromagnet heterostructure in the LTEM environment.[28] These works demonstrate the critical importance of integrating electrical current driven skyrmion devices with biasing holders that enable operando LTEM measurements.

In this work, we present a microfabricated device that enables the operando current pulsing to chiral magnetic samples in LTEM. We observe strongly pinned skyrmions that can be nucleated with electric current but are strongly bound to pinning sites. Using a thermally assisted nucleation and annihilation mechanism via current-induced Joule heating, we control the skyrmion density in these devices. These findings highlight the potential value of skyrmion pinning sites and provide an alternative to controlling skyrmion density from previously studied mechanisms such as changing the tilt angle of the magnetic field or varying the temperature.[29,30] Moreover, micromagnetic simulations clarify that the nucleation and annihilation of skyrmions are consistent with a thermally assisted process in which the spin texture finds its energetic minimum of configuration space. Finally, we study the magnetic field stability of skyrmion textures as a function of their skyrmion density and find that as the density increases, the textures become increasingly stable against magnetic field perturbations.

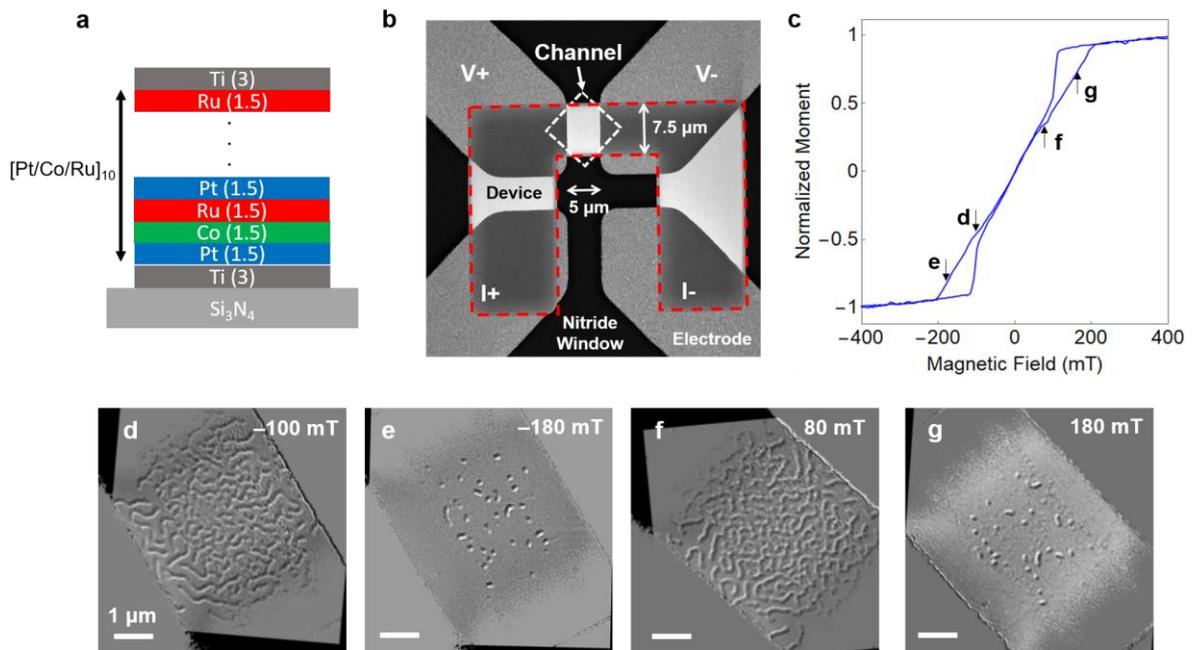

**Figure 1.** (a) Sample material composition. (b) High-angle annular dark-field image of operando LTEM device. Red dashed line defines the boundary of the device. (c) Magnetic hysteresis for a magnetic field applied out of the plane. (d-g) LTEM images from the device channel at different magnetic fields in the sequential order the magnetic field was varied.

In this study, we investigate Néel-type chiral textures stabilized in heavy metal ferromagnet heterostructures,[7,8,33,11,12,15,19,29–32] with ten repeats of a Pt(1.5 nm)/Co(1.5 nm)/Ru(1.5 nm) trilayer stack (Fig 1.a) deposited by magnetron sputtering in an Ar environment deposition chamber with a base pressure of ~$10^{-7}$ Torr. A 3 nm layer of Ti was used for seeding and capping. We grow identical films on a patterned device template for the operando experiment and two additional substrates: a thermally oxidized silicon substrate and an unpatterned $SiN_x$ membrane. A film grown on a thermally oxidized silicon substrate is used to characterize magnetic properties. At room temperature, the saturation magnetization is $M_s$=1340 kA/m, and the effective anisotropy is $K_{eff}$=56 kJ/m$^3$ as measured using vibrating sample magnetometry (VSM). The positive sign of $K_{eff}$ indicates that the sample exhibits an out-of-plane easy axis. Fig 1.c shows the out-of-plane magnetic moment normalized by $M_s$ as a function of magnetic field.

An extended film deposited on an unpatterned $SiN_x$ membrane is used to observe the magnetic behavior as a function of magnetic field. Additionally, this sample allows us to quantify the DMI constant using Lorentz microscopy. Since no contrast in the LTEM image is produced by a Néel type domain wall when the sample is normal to the beam propagation direction,[11,34] we measure with a 4 mm defocus and a 20° stage tilt. In this setting, we observe the breaking up of labyrinthine domains into skyrmions as we increase the magnetic field from zero to saturation,[30] whereas decreasing the field from saturation to zero does not lead to nucleation of skyrmions (Supplemental material S1). We find an average domain wall width of $L_0$=126 nm from the Fourier transform of labyrinthine domain state images at zero field. We calculate the DMI constant ($|D|$) from a wall energy model using $L_0 = d\frac{\pi}{2} exp\left(\frac{\pi \sigma_{DW}}{\mu_0 M_s^2 d} - \frac{1}{2}\right)$[22,32,35–37], where $d$ is the thickness of Co layer and $\sigma_{DW}$ is the domain wall energy. Assuming $A_{ex}$ = 30 pJ/m,[38] we calculate the DMI constant from $\sigma_{DW} = 4\sqrt{A_{ex} K_{eff}} - \pi|D|$ and find $|D|$=0.11 mJ/m$^2$. We also find that these parameters yield the same domain width when they are used as inputs to micromagnetic simulations. The observation of isolated skyrmions in our experiment in spite of a DMI on the lower end of literature values[39] indicates that the dipolar field is playing a significant role in formation of skyrmions in our sample, which is also consistent with a prediction for skyrmions with diameters between 10 to 100 nm.[40]

For operando experiments, we fabricate Π-shaped devices using photolithography and lift-off techniques directly on protochip™ fusion e-cell composed of a 50 nm thick $SiN_x$ membrane window and gold electrodes that extend to the center of the membrane (Fig 1.b). Each corner of the channel is in electrical contact with contact pads that connect via spring contacts to electrical equipment outside of the transmission electron microscope. This geometry allows 4-terminal sensing with the device region between the voltage leads as the active channel imaged by LTEM. To accurately determine the current density without the confounding influence of the lead and contact resistances, we first measure the channel resistance, $R = 12 \sim 14\ \Omega$, in the 4-terminal geometry. We monitor the voltage ΔV across $V_+$/$V_-$ on an oscilloscope as we apply pulses using arbitrary waveform generator through $I_+$/$I_-$. We then calculate the current density using $J = \frac{\Delta V}{RA}$, where A is the cross-sectional area of the film. Magnetic domains from the active device channel observed at different points in the hysteresis loop are shown in part d-g of Fig 1. Similar to films on the unpatterned $SiN_x$ membrane, we find the transition from labyrinthine domains (Fig 1.f) to isolated skyrmions (Fig 1.g) when increasing the field from zero to saturation but no nucleation in the reverse direction starting from saturation.

As we increase the magnetic field from zero, we notice that a higher skyrmion density forms in the operando device as compared to the unpatterned film grown on the $SiN_x$ membrane. Although the peak skyrmion density is found close to $H$=170 mT for both samples, the skyrmion density in the operando device is up to 100 times higher than in the unpatterned $SiN_x$ membrane. We also find a spatial non-uniformity of the density within the operando device. Areas close to the electrode show a smaller skyrmion density compared to the center of the channel. Further investigation of the sample reveals that the fusion e-cell chips provided from Protochips™ contain a manufacturing residue of gold nanoparticles on the SiN$_x$ surface with a diameter between 5 nm and 15 nm (supplemental material S2). The films that grow on these particles show local buckling and intermixing of the repeated multilayers, creating strong local pinning sites for magnetic textures. Within the channel area, the density of gold nanoparticles is sparser near the electrode and more concentrated towards the center of the membrane. This is reflected in the non-uniform distribution of field nucleated skyrmion between region 1 and region 2, as specified in Fig 2.a.

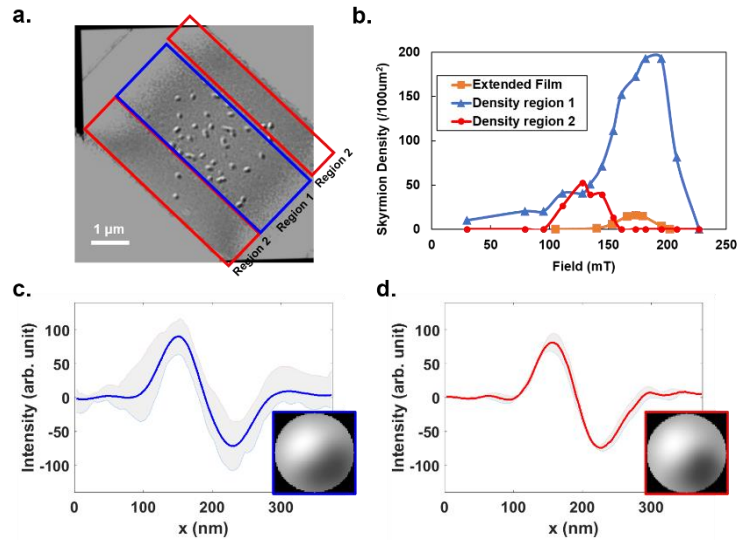

**Figure 2.** (a) LTEM image of field nucleated skyrmions in the device at H=170 mT. (b) Skyrmion density measured from region 1 and region 2 of the device, and from the extended film deposited on a SiN$_x$ membrane. (c, d) Average line cut profile of skyrmions found in region 1 and region 2. The gray area shows the maximum and minimum intensity of the profile obtained from all skyrmions found in respective regions (inset) averaged intensity map of skyrmions.

Because the field-nucleated skyrmion density is different in region 1 and region 2 of the operando device, we compare the internal structure of the skyrmions in each of the two regions using the intensity profile of skyrmions in LTEM image. The average diameter of skyrmions is $d=94 \pm 11$ nm and $d=99 \pm 10$ nm for skyrmions in region 1 and region 2, respectively, which is similar to the $d=101$ nm found from skyrmions on the film grown on an unpatterned $SiN_x$ membrane. Moreover, an averaged linecut profile and intensity map of the skyrmions obtained from region 1 (Fig 2.c) and region 2 (Fig 2.d) are indistinguishable. This suggests that although more skyrmions are nucleated due to the anchoring of the domain walls by pinning sites in region 1, the overall structure and size are not affected.

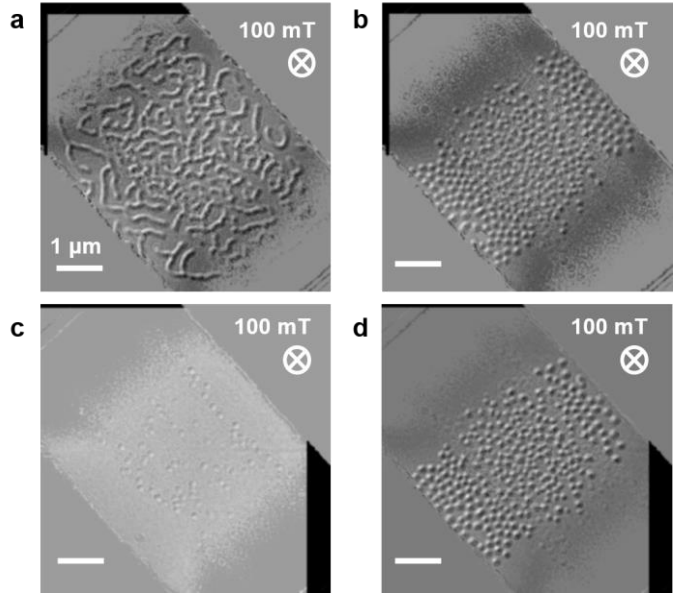

**Figure 3.** (a) A stripe domain state at magnetic field $H=100$ mT and (b) a nucleated skyrmion lattice after application of a current pulse (c) A field polarized state at the same magnetic field and (d) a nucleated skyrmion lattice after application of a current pulse. In both cases, the current pulse with current density $J=7.2\times10^{11}$ A/m$^2$ and pulse width $t=300$ ns was applied.

Next, we examine the response of chiral textures to electric current pulses. Due to limitations in bandwidth of the holder, the current pulses have a rise time of $\tau\sim100$ ns. Thus, we apply current pulses of width 50 ns or greater to minimize the deformation of the pulse shape. The pulse width presented in the remainder of this work is a nominal value and the current density is the peak current density obtained from the measured pulse shape in an oscilloscope in the 4-terminal configuration. Before applying the current, the sample is initialized to the labyrinth state by applying a strong negative field beyond saturation, passing through zero-field and then setting $H$=+100 mT (Fig 3.a). Upon application of the current pulse with current density $J$=7.2×10$^{11}$ A/m$^2$ and pulse width 300 ns, we observe the nucleation of skyrmions from labyrinthine domains (Fig 3.b). To examine if skyrmion nucleation occurs as a search for the energy minimum, we also initialize the sample with field polarized state at the same field by applying a strong positive field beyond saturation and reducing the field to $H$=+100 mT (Fig 3.c). The sample stays at a field polarized single domain state due to the flat hysteresis loop (Fig. 1c). We find skyrmions are nucleated with the similar density using the same current pulse, which supports our picture of thermally assisted energy minimization.

Recently, the role of defects in ferromagnetic multilayer films supporting skyrmions has been of key interest because the defect may contribute to the behavior of skyrmions during transportation, creation, and annihilation process.[41–43] While some studies suggest that skyrmions can move efficiently around defects in a racetrack[44] or reduce the deflection angle from skyrmion Hall effect,[14] other studies have concluded that they could be a source of pinning for domain walls and skyrmions.[15,43,45,46] Also, non-trivial collective motion and dynamics of skyrmions have been predicted, assuming Thiele's equation of motion in the presence of pinning potentials.[41,42,47] Experimentally, we find no evidence of current-induced translational motion of skyrmions in our device, which we attribute to the strong pinning of skyrmions by defects. Whether there is a contribution from more complex dynamic behaviors such as jamming or collective motion[42,47,48] is a subject of further study. Nevertheless, the immobility of the skyrmions suggests that the spin orbit-torque contributes as a secondary effect in our experiment.

To investigate how current pulses affect the density of the skyrmions, we analyze the statistics of skyrmion nucleation and annihilation under varied external magnetic fields and also using the thermal energy deposited by a current pulse as control variables. We classify three different regimes of behavior in terms of three magnetic fields that we label high field ($H$=180 mT), intermediate field ($H$=140 mT), and low field ($H$=90 mT). We calculate thermal energy via Joule heating from, $E = \int V^2(t)/R\, dt$, where the square of voltage measured from the oscilloscope (Supplemental material S3) is divided by the resistance and integrated over the duration of the current pulse.

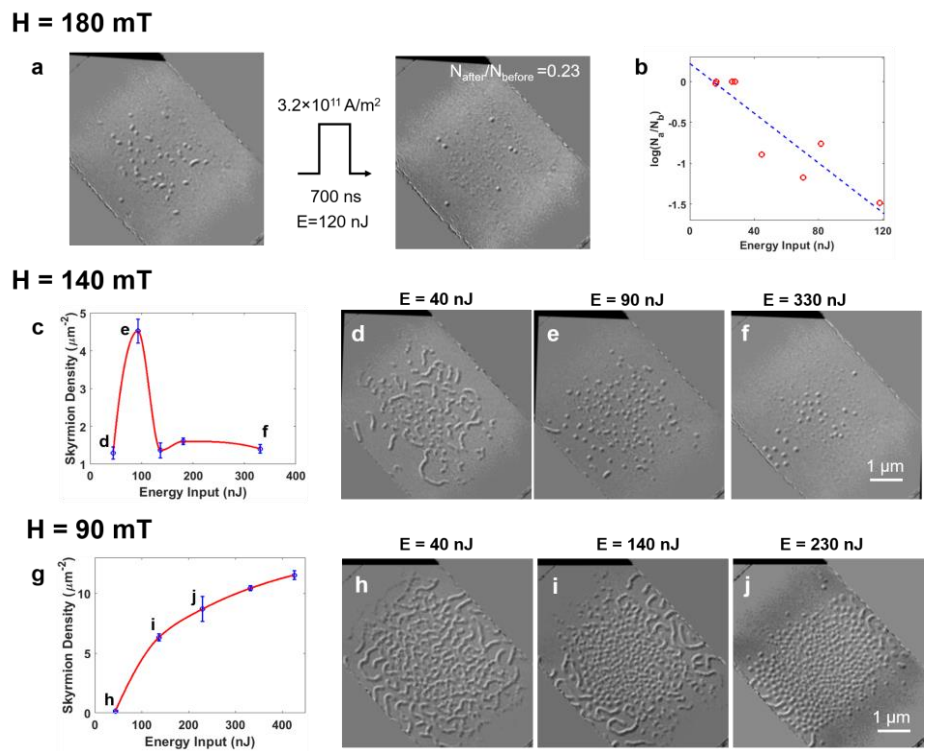

**Figure 4.** (a) Annihilation of skyrmions after a current pulse at $H$=180 mT and (b) the trend of annihilation as a function of current pulse energy. The blue dashed line is a fit to an Arrhenius function. (c), (g) The trend of skyrmion density as a function of current pulse energy at $H$=140 mT and $H$=90 mT, respectively. (d-f) and (h-j) shows the image of the channel after applying current pulse at different points in the curve from (c) and (g), respectively.

We start at high field where skyrmions with a density of 1.2±0.4 μm$^{-2}$ are field nucleated using the procedure described above. After application of a 700 ns current pulse at a $J$=3.2×10$^{11}$ A/m$^2$ current density, we observe that the number of skyrmions decreases to around a quarter of the initial number (Fig 4.a). Next, we varied the pulse width from 200 ns to 700 ns at a fixed current density of $J=3.2\times10^{11}$ A/m² to study how the number of skyrmions decrease as the pulse width increases. To see a more general trend of skyrmion annihilation as a function of pulse energy, we expanded the range of applied current density from $J=1.6\times10^{11}$ A/m² to $J=5\times10^{11}$ A/m² and pulse widths from 100 ns to 700 ns. We plot the ratio of the number of skyrmions before and after the current pulse ($N_a/N_b$) as a function of input energy (Fig 4.b). Under the assumption that the temperature increase $\Delta T$ varies linearly with the current pulse energy, the data follows an Arrhenius trend as indicated by the fit line. While this result does not elucidate a detailed mechanism of annihilation, it supports a thermally assisted process in which the spin texture is excited by current pulses, likely including Joule heating, to leave a local energy minimum and find a more stable energy minimum in configuration space.

At intermediate and low magnetic fields in which the magnetic initialization leads to a labyrinth state, skyrmions nucleate by breaking apart the labyrinthine domains during the current pulse as we observed earlier (Fig 3). Because the initial states contain zero or few skyrmions, we only consider the density of skyrmions after current pulses for the analysis in Fig 4.c and Fig 4.g. When current pulses are applied at intermediate field ($H=140$ mT), we observe an initial increase in skyrmion density to 4.5±0.3 μm⁻² using $E=90$ nJ pulses. The density then drops and stabilizes to ~1.5 μm⁻² for the higher energy pulses. We note that a mixed state of isolated skyrmions and labyrinthine domains appears before the skyrmion density reaches a maximum, whereas only isolated skyrmions are present when the density saturates to a lower value. This suggests that the skyrmions are first nucleated by the breaking apart of labyrinthine domains via a thermally assisted process but are still metastable. When the injected thermal energy exceeds a threshold value, it contributes to skyrmion annihilation that drives the system toward a more stable energy minimum in the skyrmion configuration state with a lower skyrmion density.

Finally, we repeat the experiment at the lowest magnetic field, *H*=90 mT. Here, the initial state is a dense labyrinthine domain state. Unlike at the intermediate field in which we observe a peak and then saturation at a lower density, at this field, we find a monotonically increasing trend of skyrmion density over a wide range of injected energy that appears to approach saturation for the largest energy pulses (Fig 4.g). We obtain the highest skyrmion density observed in the study, 11.5±0.3 μm$^{-2}$, which is more than five times higher than what we obtained from field nucleation. Moreover, the monotonic trend of the skyrmion density with respect to injected energy indicates that we can achieve arbitrary skyrmion density between 0 and 11.5 μm$^{-2}$ by applying an appropriate current pulse to the device.

To obtain micromagnetic insights into the experimental results, we carry out a simulation using MuMax3.[49] The simulation was performed on a 1 × 1 μm$^2$ area with a cell size of 4 × 4 × 1.5 nm$^3$. All material parameters are the same as those determined for the experimentally studied films. The total magnetic field is applied, as in the experiment, at a sample tilt angle of 20°. We assume a periodic boundary condition in the transverse direction to disregard edge effects and in the perpendicular direction to simulate the repeats of ferromagnetic thin film spaced by non-magnetic layers.

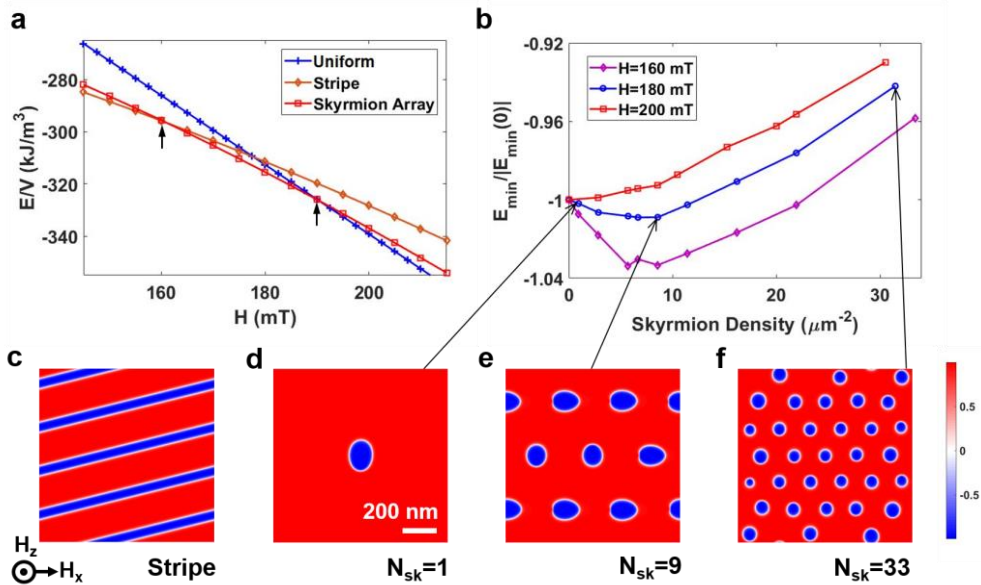

**Figure 5.** (a) Energy density of uniform, stripe domain and skyrmion array states as a function of magnetic field. Arrows at 160 mT and 190 mT indicate the transition fields. (b) Minimized total energy from the simulation of skyrmion array as a function of skyrmion density normalized to the energy when skyrmion density is zero. Simulated magnetic textures on the bottom show the simulated magnetic texture for stripe domain at H=180 mT (c) and at points designated along the curve H=180 mT (d-f). The magnetic field is applied with a tilt angle of 20 degrees from the z-axis.

We first investigate the energetics of spin textures, searching for the texture that yields the lowest energy at different magnetic fields. To obtain the total energy of a magnetic texture, we initialize the simulation with each texture at zero field and evolve it over an increasing field, minimizing the total energy at a regular field interval. Skyrmion arrays for the field-dependent simulation are initialized with $N_{sk}$=9 (Fig 5.e), which was preserved over the entire range of the field we apply in Fig 5.a. Although this approach does not reveal the full energy landscape including the magnitude of the potential barrier for skyrmion annihilation, it gives us a qualitative picture of the system ground state. Thermal excitation of the sample will enable the magnetic textures to leave local energy minima and, in some cases, enter the ground state.

We observe two transitions of the ground states within the range of the field we apply. At magnetic fields below 160 mT (left arrow in Fig. 5.a), stripe domains have the lowest energy, which explains their prominence at low magnetic fields in the experiment. Above 160 mT, a skyrmion array becomes the most energetically favorable state until the second transition at 190 mT (right arrow in Fig. 5.a) above which the field polarized uniform state has the lowest energy. This observation is in agreement with Ref. [50] that used a similar computational approach.

To further illuminate the relationship between skyrmion density and the magnetic field, we calculate the system energy as we vary the skyrmion density around the magnetic fields in which a skyrmion array is the ground state. Parts (d-f) of Fig 5 show the simulated spin textures with different skyrmion densities after the system is relaxed. We record the total energy of each of these states and plot it as a function of skyrmion density in Fig 5.b. At field $H$=200 mT, the global energy minimum is a ferromagnetic state where skyrmion density is zero. As we lower the applied magnetic field to $H$=180 mT, an energy minimum emerges at non-zero skyrmion density. The energy minimum becomes more pronounced as the field further decreases to $H$=160 mT. This simulation result is consistent with our experimental results in which we observe skyrmion annihilation at high fields and an increasing trend of skyrmion density after nucleation at lower fields.

When skyrmions are closely packed, the interaction between skyrmions can dominate the stability of skyrmions making the annihilation process different from that of isolated skyrmions.[22,51] To verify this idea, we experimentally prepare our device using current pulse nucleation, choosing magnetic fields such that we systematically vary the skyrmion density.

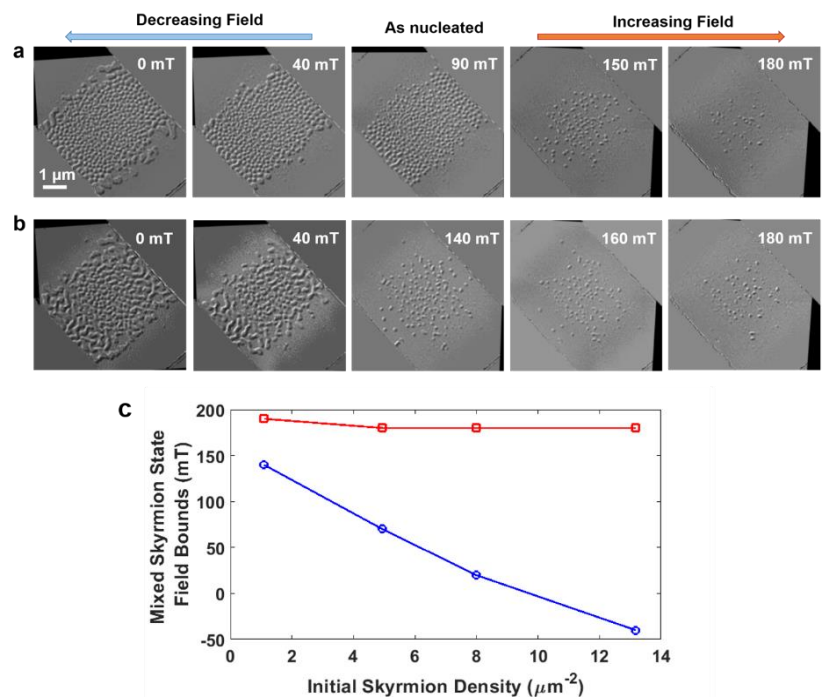

**Figure 6.** Stability of nucleated skyrmions to the changes in the magnetic field. Skyrmions are nucleated at (a) 90 mT and (b) 140 mT and subject to changes in the field. After the nucleation, images were taken sequentially as the field increases or decreases. Reinitialization is done only when the direction of the field variation changes. (c) Upper bound (red squares) and lower bound (blue circles) of the magnetic field at which skyrmionic bubbles are found as a function of initial skyrmion density.

We then increase (or decrease) the magnetic field to find the magnetic field at which the skyrmions are annihilated, thus establishing the magnetic field range over which skyrmions persist. The evolution of spin textures follow different paths depending on the direction of the field variation. When the field increases, the skyrmions collapse and the density decreases until we are left with a field polarized magnetic state (orange arrow, Fig 6.a, b). For increasing fields, the behavior is independent of the skyrmion density. On the other hand, when the field decreases, the skyrmions expand and in some cases begin to form short labyrinthine domains (blue arrow, Fig 6.a, b). As the chiral textures expand, however, adjacent domain walls repel each other,[41,52] which preserves the roughly circular skyrmion morphology. Therefore, as the initial density of skyrmion increases, individual skyrmions face a barrier to expansion into a labyrinthine domain. As a result, while the upper bound of the field that supports skyrmions remain consistent, the lower bound of the field that preserves the morphology of skyrmions decreases with the initial skyrmion density (Fig 6.c). While a similar effect has been observed in micrometer-scale skyrmions thermally nucleated by lasers,[22] we demonstrate that this type of topological protection is more general, independent of skyrmion size and the material system supporting it. It is also to be noted that similar expansion of skyrmions into short labyrinthine domains has also observed for sub 10-nm size skyrmions.[53]

In conclusion, we demonstrate a device for high-resolution, operando LTEM current pulsing experiments. We find that skyrmions can be robustly generated in samples with strong pinning sites. While the details of skyrmion nucleation and annihilation are determined by the interplay of several parameters, the density of skyrmions can be controlled with current-induced thermal pulses. Using this effect, we initialize our device with controlled skyrmion densities and verify an enhanced stability of high-density skyrmions that is persistent over a wide range of magnetic fields. Our work on thermally assisted control of skyrmion density in films with strong pinning suggests a new possibility to engineer skyrmion density and dynamics using defects that are inherent to the device.[54–56] Also, the experimental framework for operando LTEM imaging that we develop offers a valuable tool for the further investigation of ultra-small skyrmions in emerging skyrmionic materials, including ferrimagnets and synthetic antiferromagnets that require truly nanoscale resolution to visualize magnetic texture and dynamics.[57,58]

DATA AVAILIBILITY

The data that support the findings of this study are available from the corresponding author upon reasonable request.

ACKNOWLEDGMENT

This work is primarily supported by the DARPA TEE program (D18AC00009). Any opinions, findings, and conclusions or recommendations expressed in this publication are those of the author(s) and do not necessarily reflect the views of DARPA. L.Z. acknowledges support by the Office of Naval Research (N00014-15-1-2449). We acknowledge the use of shared facilities including the Cornell Center for Materials research, and NSF MRSEC (DMR-1719875) and the Cornell NanoScale Facility, an NNCI member supported by NSF (NNCI-1542081). We thank Prof. Robert A. Buhrman for helpful discussion.

Supplemental Material

# Operando Control of Skyrmion Density in a Lorentz Transmission Electron Microscope Using Joule Heating

*Albert M. Park, Zhen Chen, Xiyue S. Zhang, Lijun Zhu, David A. Muller, Gregory D. Fuchs\**

S1. Ferromagnetic multilayer grown on an unpatterned SiN$_x$ membrane

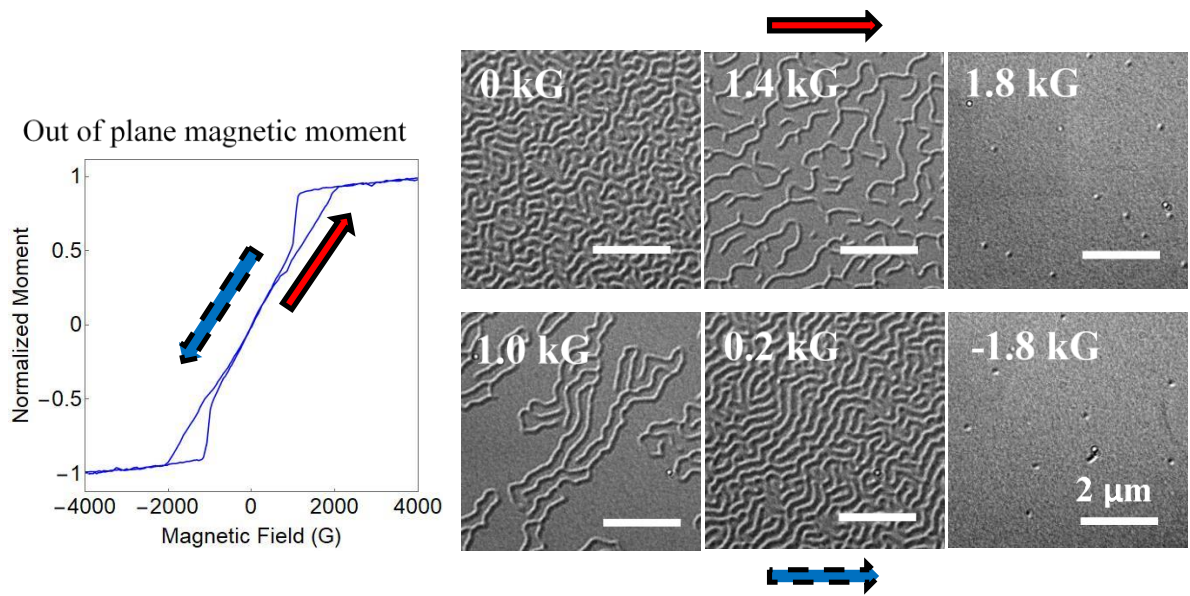

**Figure S1**. Out-of-plane magnetic hysteresis and LTEM images of spin texture at various magnetic fields along the increasing (red arrow) and decreasing (blue arrow) field for the film grown on an unpatterned SiN$_x$ membrane

S2. Gold nanoparticles on Protochip$^{TM}$ e-cell membrane

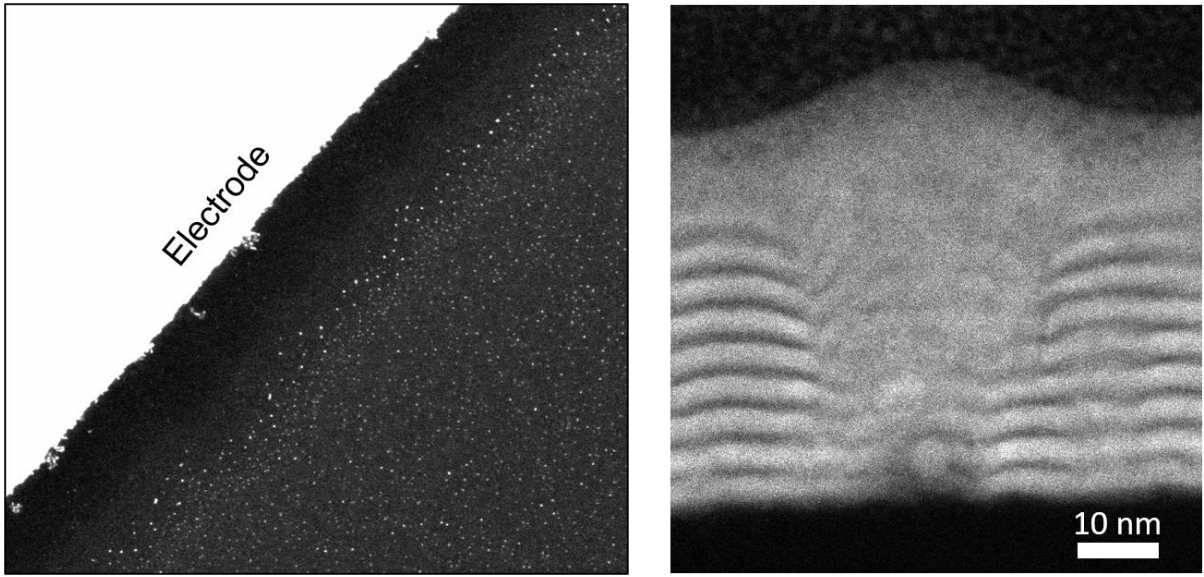

**Figure S2.** Plane view TEM image of gold nanoparticles on as received Protochip$^{TM}$ e-cell membrane and the multilayer mixing and the film buckling observed on the site of a

S3. Current pulse shape from the oscilloscope

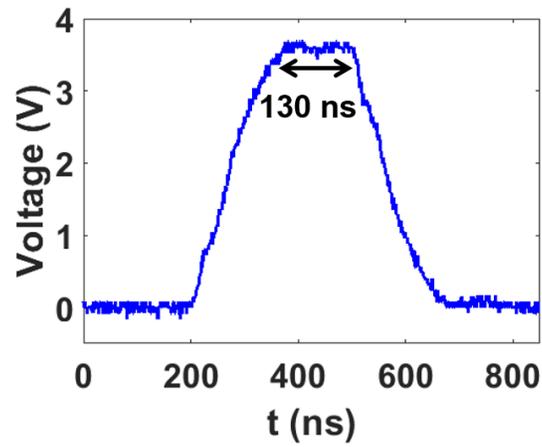

**Figure S3.** Applied current pulse measured from the oscilloscope for the 300 ns nominal pulse width. Rise time of $\tau\sim100$ ns is observed due to bandwidth of the holder.